\documentclass[3p,times,procedia]{elsarticle}
\usepackage{nupha_ecrc}

\volume{00}

\firstpage{1}

\journalname{Nuclear Physics A}

\runauth{}

\jid{nupha}

\jnltitlelogo{Nuclear Physics A}

\usepackage{amssymb}

\usepackage[figuresright]{rotating}

\begin{document}

\begin{frontmatter}



\dochead{XXVIIIth International Conference on Ultrarelativistic Nucleus-Nucleus Collisions\\ (Quark Matter 2019)}

\title{How EIC can help us to understand heavy-ion collisions }


\author{Yoshitaka Hatta}

\address{Brookhaven National Laboratory, Upton, NY 11973, USA}

\begin{abstract}
I give a brief overview of the science cases of the Electron-Ion Collider (EIC)  with a particular emphasis on the connections to the physics of ultrarelativistic heavy-ion collisions.  
\end{abstract}

\begin{keyword}
EIC \sep heavy-ion collisions \sep  tomography \sep proton spin \sep gluon saturation

\end{keyword}

\end{frontmatter}


\section{Introduction}
\label{}

This talk is meant to be an brief introduction to the science cases of the Electron-Ion Collider (EIC) for the researchers in ultrarelativistic heavy-ion collisions.  EIC is a future $ep$ and $eA$ collider dedicated to the study of the nucleon and nucleus structures, to be built at Brookhaven National Laboratory (BNL) in the U.S. The first collision is expected to take place  about 10 years from now, so it may seem like a distant future. However, there is already a large user community with more than 1000 members from over 200 institutions (visit the website of the user group www.eicug.org). Given the recent `Critical Decision-0' (CD0, an official start of the project) by the US Department of Energy, I feel that now is the perfect time to reemphasize and further explore the possible connections between the physics of EIC and heavy-ion collisions. 

Let me first list the basic facts and numbers of EIC. The experiment at EIC is Deep Inelastic Scattering (DIS) off a proton or a nucleus with the variable center-of-mass energy within the range $20 <\sqrt{s}<140$ GeV, with the design luminosity  $\sim$10$^{34}$ cm$^{-2}$ s$^{-1}$. The energy is somewhat lower than at HERA ($\sqrt{s}=318$ GeV), the previous $ep$ collider at DESY,  but the luminosity is higher by a factor of 1000. EIC is the world's first polarized $ep$ collider. The existing polarized $ep$ experiments (HERMES at HERA and COMPASS at CERN) are fixed-target ones. Switching to a collider makes a huge difference in kinematical coverage, about 2 orders of magnitude in $x$ and $Q^2$. EIC is also the first high energy $eA$ collider, again with an unprecedented coverage in kinematics. The target is not just proton, but it can be  deuterium, helium, carbon, uranium...any nucleus of your choice, and light nuclei can be polarized.  These characteristics make EIC a unique and versatile machine to explore  hitherto untouched landscapes of QCD. I should also mention other current and future $ep$ experiments in the world. Jefferson laboratory (JLab) has been conducting low energy, fixed-target $ep$ and $eA$  experiments with  a 12 GeV electron beam.  In China, a future $ep$ and $eA$ collider (EIcC) is planned at a lower center of mass energy $\sqrt{s}\lesssim 30$ GeV which is in many ways complementary to the U.S. EIC. Also, at the LHC, there are plans for future $ep$ experiments (LHeC, VHEeP, FCC-eh) in the TeV energy region. All these suggest that nucleon structure studies in DIS will continue to be one of the dominant trends in nuclear physics and QCD in the foreseeable future.

In the following, I will give a brief review of each of the main physics cases to be explored at  EIC, with a particular emphasis on the topics that may be of interest to heavy-ion physicists. The list is by no means complete, and I refer to the EIC white paper \cite{Accardi:2012qut}, 
 and the recent INT workshop proceedings \cite{Aidala:2020mzt} for more details.

\section{Nucleon tomography}

Tomography is an important key word in EIC physics. It is basically a technique to see inside an object without cutting it, like a CT (computed tomography) scan for cancer treatment. The object of interest for us is the nucleons or nuclei, and we would like to understand how the partons (quarks and gluons) are distributed inside them. By distribution I mean a multi-dimensional one. The familiar Parton Distribution Functions (PDFs) $u(x),d(x),...$, where $x=E_{parton}/E_{hadron}$ is the longitudinal momentum fraction of the parent hadron carried by $u,d,...$ quarks, is a  one-dimensional object from our point of view.  In reality, partons have transverse momentum $k_\perp$ and are also distributed in impact parameter space $b_\perp$. One can thus generalize the notion of PDF by including the dependence on these parameters. This leads to the transverse momentum dependent (TMD) distributions $u(x,k_\perp)$, and the generalized parton distributions (GPD) $u(x,b_\perp)$ and ultimately the Wigner distribution $W(x,k_\perp,b_\perp)$ which is a `phase space' distribution. Importantly, these distributions are not only of conceptual interest, but they  can be practically used in describing observables. 

\subsection{TMD} 
For TMDs, a rigorous  framework based on QCD factorization has been known since the 80s, and this allows one to compute, for example, the $P_\perp$ spectrum in semi-inclusive DIS (SIDIS) $ep\to e'h(P_\perp)X$ at low transverse momentum $P_\perp \sim \Lambda_{QCD}$. Schematically, the formula reads
\begin{eqnarray}
\frac{d\sigma}{dP_\perp}= H(\mu)\int d^2q_\perp d^2k_\perp f(x,k_\perp,\mu,\zeta)D(z,q_\perp,\mu,Q^2/\zeta) \delta(zk_\perp+q_\perp-P_\perp)+\cdots
\end{eqnarray}
where $f$ and $D$ are the TMD versions of the PDF and the fragmentation function. Behind this intuitive formula,  enormous complications and subtleties are hidden. To say the least, note that TMDs depend on {\it two} renormalization scales $\mu,\zeta$ instead of one for PDF.  
The  advancement in theory in terms of higher order calculations over the past several years has been really remarkable, and nowadays people have started doing a global analysis \cite{Scimemi:2019cmh,Bacchetta:2019sam}. At the moment, the available data points are still very much limited compared to those in global QCD analyses for the collinear PDFs, but this will change dramatically in the coming EIC era. The outcome of such a global analysis will be a vivid 3-dimensional snapshot of the proton wavefunction in momentum space. 

The $k_\perp$-dependent distributions have been routinely used in the heavy-ion community, especially among theorists working on Color Glass Condensate (CGC), although they are not usually called `TMD', but instead `unintegrated gluon distribution'.  Usually only the gluon distribution at small-$x$ is considered because of an overwhelming emphasis on  the role of small-$x$ gluons in heavy-ion collisions.  The difference appears to be more than  just names. These two types $k_\perp$-dependent distributions have been studied in different frameworks  by theorists from  different communities without much interactions.  But attempts to bridge the gap between the two formalisms do exist \cite{Balitsky:2016dgz}. I think such efforts will gain more importance in future.  
 
 \subsection{GPD}

GPDs can be accessed in deeply virtual Compton scattering (DVCS) $ep \to e'\gamma^*p\to e'\gamma p'$. At high energy,  the momentum transfer $t=(p-p')^2$ of the elastically scattered proton is dominated by the transverse component $t\approx -\Delta_\perp^2$. This is Fourier conjugate to  the impact parameter $b_\perp$. The measurement of GPDs thus tells us how the partons are spread in impact parameter space at a given value of $x$.
An important theme  in the GPD community of late is  the so-called gravitational form factors defined as the off-forward matrix element of the QCD energy momentum tensor $T_{\mu\nu}^{q,g}$ (The subscript $q,g$ refers to the quark/gluon part.)
\begin{eqnarray}
\langle p'|T_{q,g}^{\mu\nu}|p\rangle = \bar{u}(p')\left[A_{q,g}(t) \gamma^{(\mu}\bar{P}^{\nu)} + B_{q,g}(t)\frac{\bar{P}^{(\mu}i\sigma^{\nu)\alpha}\Delta_\alpha}{2M} +D_{q,g}(t) \frac{\Delta^\mu \Delta^\nu-g^{\mu\nu}\Delta^2}{4M} + \bar{C}_{q,g}(t) Mg^{\mu\nu} \right]u(p) 
\end{eqnarray}
These form factors describe how the proton couples to a graviton. Of course, they cannot be measured directly because the gravitational interaction  is too weak. But they can be measured indirectly as certain moments of the GPDs. The $A,B$ form factors have been the primary motivation of  GPD studies due to their connection to the Ji sum rule for the proton spin \cite{Ji:1996ek}.  Recently,  the $D$-form factor has received a lot of attention. The value of the total  $D=D_q+D_g$ at zero momentum transfer is a fundamental constant in Nature, just like the mass and spin of the proton. Moreover, after Fourier transforming to the coordinate space, it can be interpreted as the radial `pressure' distribution inside the proton, see the first extraction from the DVCS data \cite{Burkert:2018bqq}.

In DVCS, it is extremely challenging to access the gluon GPDs, even at the EIC, and here I can see possible connections to heavy-ion physics. Indeed, the distribution of gluons inside a proton (or even in a nucleus) in impact parameter space  has been  discussed by  heavy-ion physicists, although it is not called `GPD'. At small-$x$, realistic  simulations of the $b_\perp$-dependence, including the BFKL    \cite{Bierlich:2019wld} and nonlinear  
\cite{Schlichting:2014ipa} QCD evolutions exist. Moreover, these approaches  can reveal not only the (average) spatial distribution,  but also more advanced information such as the event-by-event fluctuations and correlations 
which are crucial to understand the collective phenomena in  heavy-ion collisions. It would be interesting if such developments can be redirected to provide some guidance to the GPD studies at EIC.   

\subsection{Wigner}

The Wigner distribution $W(x,k_\perp,b_\perp)$, often called `Mother distribution', contains more information about the nucleon structure than TMD and GPD combined, but compared to these lower dimensional counterparts, it has been much less studied/understood. For a long time  it was believed that the Wigner distribution was simply  impossible to measure in experiments. However, it turns out that the small-$x$ community has been routinely using it for a long time,  without calling it `Wigner'. At small-$x$, the gluon Wigner distribution is proportional to the so-called dipole S-matrix which is a fundamental object in the gluon saturation physics. Based on this, an concrete experimental observable at EIC that can access the Wigner distribution has been proposed   \cite{Hatta:2016dxp} see also \cite{Altinoluk:2015dpi,Mantysaari:2019csc}. In fact, the same observable---coherent diffractive dijet production---can be studied already at RHIC in ultraperipheral $pA$ collisions  \cite{Hagiwara:2017fye} where the equivalent photons from the nucleus can mimic the virtual photon in DIS. This can be done in parallel with the  $pA$ programs in heavy-ion physics.

\section{Proton spin}

Spin is an essential part of EIC physics. One of the most obvious and achievable goals of EIC is  to constrain the value of  $\Delta G$, the gluon helicity contribution to the proton spin   in the Jaffe-Manohar sum rule 
\begin{eqnarray}
\frac{1}{2}=\frac{1}{2}\Delta\Sigma + \Delta G + L_q+L_g, \label{jm}
\end{eqnarray}
where $\Delta \Sigma$ represents the quark helicity and $L_{q,g}$ are the orbital angular momentums of quarks and gluons.   
$\Delta G$ is given by the integral of the polarized gluon distribution $\Delta G=\int_0^1 dx \Delta G(x)$. After a decade of experimental efforts at RHIC and other facilities, the contribution to $\Delta G$ from the large-$x$ region is relatively well constrained, but there are huge uncertainties remaining in the small-$x$ region ($x<0.05$) \cite{deFlorian:2019zkl}. This will be settled down at the EIC. In the meantime, theorists are revisiting the small-$x$ behavior of the polarized parton distributions \cite{Kovchegov:2016zex}. On the other hand, measuring  the orbital angular momentum $L_{q,g}$ is  quite challenging, but EIC should seriously address this question in order to fully understand the proton spin structure.  

Any connection to heavy-ion physics? I always thought that spin was the most distant subject from heavy-ions.   But I was surprised that in this conference there are many talks on global and local  polarizations, the angular momentum generated in non-central collisions and its transfer to the polarization of measured hadrons. Some of the discussions (like `canonical' vs. `kinetic' angular momentum) are familiar to the experts of QCD spin.  In particular, the phase space Wigner distribution has been used to describe the polarization phenomenon \cite{Becattini:2013fla,Fang:2016vpj}. In this respect I would like to point out that a rigorous definition of the partonic orbital angular momentum $L_{q,g}$ in (\ref{jm}) also involves the Wigner distribution mentioned in the previous section  \cite{Lorce:2011kd,Hatta:2011ku,Ji:2012sj}. Thus it may be possible for the two communities to benefit from each other.

The decomposition (\ref{jm}) is for a longitudinally polarized proton. There are many interesting topics for a transversely polarized proton. Especially the origin of single spin asymmetry (SSA),  the left-right asymmetry of the produced hadrons with respect to the spin axis, is not yet fully understood. This will be extensively studied   at the EIC. One possible connection to the physics of heavy-ion collisions is that  SSA has also been measured at RHIC in $pA$ collisions by the STAR and PHENIX collaborations \cite{Aidala:2019ctp,Dilks:2016ufy}.  The dependence of SSA on the atomic number $A$, caused by the saturation effect in the nucleus, may help to disentangle different mechanisms of SSA \cite{Hatta:2016khv}. 

\section{Jets}

EIC is also a unique laboratory to study certain aspects of jets and jet quenching  \cite{Page:2019gbf}. Because the energy is not very high compared to the LHC, assumptions  which can be taken for granted at the LHC (such as the separation of  scales) may not work  at the EIC.  On the other hand, one expects less pileups and underlying events, so EIC can provide novel  opportunities to study power corrections and nonperturbative effects in a cleaner environment. Precision pQCD calculations can also be done, as in the recent   next-to-next-to leading order (NNLO) prediction \cite{Abelof:2016pby} for inclusive jet cross section at the EIC.  

Of course, for heavy-ion physicists, the most interesting aspect regarding jets at EIC is jet quenching in a cold nuclear matter. An experience with $pA$ collisions at the LHC \cite{Adam:2015hoa} suggests that  quenching is not very strong, but this may help to better  discriminate different approaches to parton energy loss if the data are accurate enough. Besides, the effect could be enhanced by employing heavy flavor as a probe, depending on different scenarios of hadronization \cite{Vitev:2019zau}. Thus, jet quenching and energy transport at EIC can produce important  insights and feedbacks to the physics of heavy-ion collisions and QGP. Yet,  there have been remarkably few recent predictions for EIC other than \cite{Vitev:2019zau}.  A related discussion of $P_T$-broadening may be found in  \cite{Liu:2018trl}.   

\section{Gluon saturation}

The gluon saturation  is arguably  the most relevant topic at EIC to the heavy-ion community. Needless to say, $eA$ collisions are  the ideal setup to study the gluon saturation.  This has been amply covered by the  plenary speakers of the previous Quark Matter conferences (see for example talks by T. Lappi at QM2009, A. Stasto at QM2011, T. Ullrich at QM2014, E. Sichtermann at QM2015,  B. Xiao at QM2017), so I do not dwell on it. Instead, let me just say that, in my perspective,   EIC can address the following fundamentally important question: {\it Can saturation become precision science?} Many of the observables that have been computed in the saturation framework are leading order results, often including part of the higher order corrections such as the running coupling effect. 
As a matter of fact, at present there is no all-order proof of  factorization  with gluon saturation in the usual pQCD sense. This is  because already the  `leading order' result contains infinitely many higher twist contributions, and QCD factorization beyond leading twist is  notoriously difficult. What one can do, however, is to check factorization order by order, by calculating   the next-leading-order (NLO)  correction to start with and demonstrate that all the divergences encountered can be absorbed into the renormalization of various distributions involved. The result is to be combined with the next-to-leading logarithmic Balitsky-Kovchegov (NLL BK)  equation \cite{Balitsky:2008zza}.  In the past several years, an impressive progress in this direction has been made  \cite{Chirilli:2011km,Boussarie:2016ogo,Boussarie:2016bkq,Roy:2019hwr}, which led  me to believe that NLO + NLL (plus possible `collinear improvement' \cite{Ducloue:2019ezk}) will be the norm of the saturation-based calculations in the EIC era. It may then be possible to perform the NLO `global analysis` of the dipole S-matrix, similarly to what was done in \cite{Albacete:2009fh}. Once these  higher-order calculations have been successfully tested at EIC, they can be applied to heavy-ion collisions with more confidence and accuracy.

\section{Proton mass}

Finally, I come to the proton mass problem. This is a relatively new topic in the context of EIC, in the sense that it was not emphasized in the  white paper \cite{Accardi:2012qut}. But in a recent report by the National Academy of Sciences  (https://doi.org/10.17226/25171), it has been identified as one of the most important problems to be addressed at EIC.  To put it  simply (but naively), the issue is that only 1\% of the proton mass is attributed to the sum of the $u,d$ current quark masses, while the origin of the remaining 99\% is unexplained.  Of course, this is a deep question which undoubtedly has to do with confinement and chiral symmetry breaking. Can a collider experiment shed any light on it? 

One way to understand the origin of the proton mass is to  decompose it into various building blocks, which can be done at the level of operators \cite{Ji:1994av}.  Similarly to the proton spin sum rule (\ref{jm}), one can write  
\begin{eqnarray}
M=M_{kin}^q+M_{kin}^g+M_a+M_m. \label{de}
\end{eqnarray}
The kinetic energy of quarks and gluons $M_{kin}^{q,g}$ are indeed measurable  as the second moment of the PDFs. The current quark mass term $M_m$ is related to the so-called nucleon sigma term. The most interesting entry in (\ref{de}) is then the  trace anomaly, or the gluon condensate contribution $M_a\sim \langle p|F^2|p\rangle$. 

It has been demonstrated  \cite{Kharzeev:1998bz,Hatta:2018ina} that $J/\psi$ photoproduction in $ep$ scattering near threshold is a promising observable to access $M_a$. Experiments are ongoing at JLab \cite{Ali:2019lzf}. $J/\psi$, because a heavy quarkonium interacts with the proton only via  gluon exchanges. Near-threshold, because in order to be sensitive to the twist-{\it four} operator $F^2$, the $\gamma^*p$ center-of-mass energy has to be as low as possible, or else the process is dominated by the twist-two contribution. However, this does not necessarily mean that the  $ep$ center-of-mass energy is small, and the process can also be studied at EIC \cite{Lomnitz:2018juf} where $\Upsilon$ production can  be measured. In fact, we can do it also at RHIC, in ultraperipheral $pA$ collisions (UPCs) in which the nucleus merely acts as a source of on-shell photons \cite{Hatta:2019lxo}. The challenge is that one has to detect $J/\psi$ and $\Upsilon$ in the very forward, low-$P_\perp$ region. This may be possible after the completion of the STAR forward upgrade.

\section{Conclusions}

Hopefully I have convinced the reader that there is a strong overlap between the EIC and heavy-ion sciences. This is not limited to the gluon saturation and jet quenching as is usually thought. 
I have explained potential feedbacks to the heavy-ion community, but at the same time, I tried to emphasize that the direction of the arrow can be reversed. Namely, the heavy-ion community can help to understand EIC physics in many ways, in particular through ultraperipheral $pA$ collisions.  

\section*{Acknowledgements}

This work is supported by the U.S. Department of Energy, Office of
Science, Office of Nuclear Physics, under contract No. DE- SC0012704,
and in part by Laboratory Directed Research and Development (LDRD)
funds from Brookhaven Science Associates.





\bibliographystyle{elsarticle-num}
\bibliography{references}







\end{document}